\documentclass{PoS}
\usepackage{graphicx}
\usepackage{siunitx}
\DeclareSIUnit\parsec{pc}

\title{A multi-PMT Optical Module for the IceCube Upgrade}

\ShortTitle{A multi-PMT Optical Module for the IceCube Upgrade}

\author{
The IceCube Collaboration\footnote{For collaboration list, see PoS(ICRC2019) 1177.}, C.\ Dorn$^\ddag$, M.\ Kossatz$^\S$, A.\ Kretzschmann$^\S$, H.-W.\ Ortjohann$^{\#}$, J.\ Reubelt$^\ddag$, K.-H.\ Sulanke$^\S$ and R.\ Weigel$^\ddag$\\

{\itshape \href{http://icecube.wisc.edu/collaboration/authors/icrc19_icecube}{http://icecube.wisc.edu/collaboration/authors/icrc19\_icecube}}\\
{$^{\ddag}$ \itshape Friedrich-Alexander Universit\"at Erlangen-N\"urnberg, Erlangen, Germany}\\
{$^{\S}$ \itshape DESY, Zeuthen, Germany}\\
{$^{\#}$ \itshape Westf\"alische Wilhelms-Universit\"at M\"unster, M\"unster, Germany}\\
E-mail:\\ \email{lew.classen@icecube.wisc.edu,\\alexander.kappes@icecube.wisc.edu,\\ timo.karg@desy.de}
}

\abstract{

Following the first observation of an astrophysical high-energy neutrino flux with the IceCube Neutrino Observatory in 2013 and the identification of a first cosmic high-energy neutrino source in 2017, the detector will be upgraded with about 700 new advanced optical sensors. This will expand IceCube's capabilities both at low and high neutrino energies. A large fraction of the upgrade modules will be multi-PMT Digital Optical Modules, mDOMs, each featuring 24 three-inch class photomultiplier tubes (PMTs) pointing uniformly in all directions, thereby providing an almost homogeneous angular coverage. The signal from each PMT is digitized individually, providing directional information for the incident photons. Together, the 24 PMTs provide an effective photosensitive area more than twice than that of the current IceCube optical module. The main mDOM design challenges arise from the constraints on the module size and power needed for the 24-channel high-voltage and readout systems. This contribution presents an mDOM design that meets these challenges and discusses the sensitivities expected from these modules.\\

\vspace{4mm}
{\bfseries Corresponding authors:}
\speaker{Lew Classen}$^{1}$, Alexander Kappes$^{1}$, Timo Karg$^{2}$\\
{$^{1}$ \itshape Westf\"alische Wilhelms-Universit\"at M\"unster, M\"unster, Germany}\\
{$^{2}$ \itshape DESY, Zeuthen, Germany}

}

\FullConference{36th International Cosmic Ray Conference -ICRC2019-\\
		July 24th - August 1st, 2019\\
		Madison, WI, U.S.A.}

\begin{document}

\section{The multi-PMT Digital Optical Module (mDOM)}
\label{sec:mdom}
\begin{figure}
\begin{minipage}{0.55\textwidth}
  \centering
    \includegraphics[trim=0cm 8.5cm 0cm 18cm, clip, width=\textwidth]{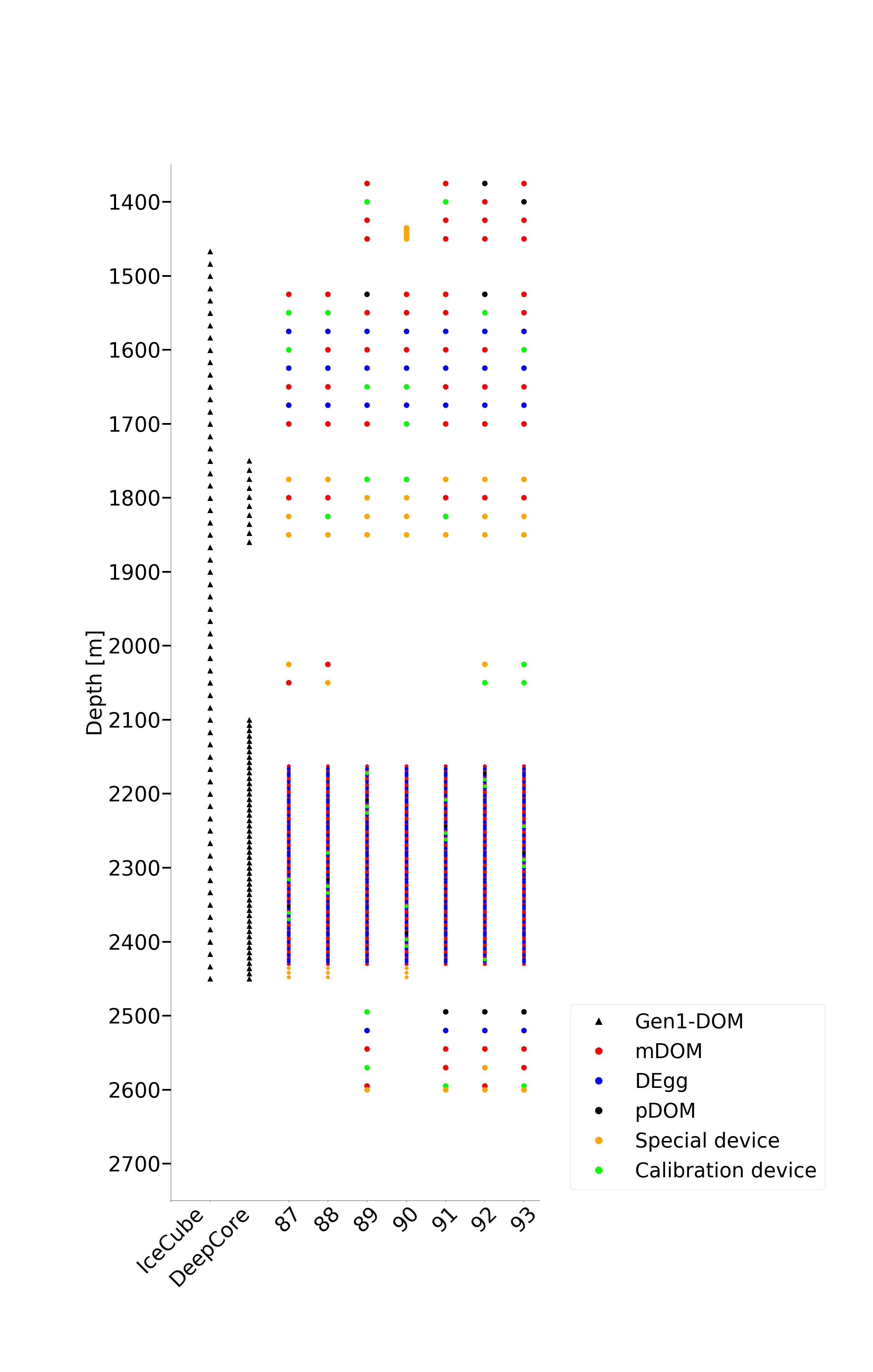}
\end{minipage}
\hfill
\begin{minipage}{0.45\textwidth}
  \centering
    \caption{Currently considered IceCube Upgrade string configurations in comparison to the strings deployed in IceCube and the DeepCore extension. Each marker represents an optical module.}
    \label{fig:icupgrade}
\end{minipage}
\end{figure}

Located in the deep glacial ice of Antarctica, IceCube \cite{Aartsen2016:icecube} is the world's largest neutrino telescope. Originally designed for the investigation of the neutrino sky on the $\si{\tera\electronvolt}$ to $\si{\peta\electronvolt}$ energy scale and beyond, the energy threshold was lowered to $\SI{\sim 10}{\giga\electronvolt}$ by the DeepCore \cite{Abbasi2012:deepcore} extension. In a next step, IceCube's capabilities will be further enhanced through the installation of about 700 new optical modules distributed over seven vertical strings (see Fig.~\ref{fig:icupgrade}), mainly located in the DeepCore region (IceCube Upgrade). This upgrade will lower IceCube's energy threshold to a few $\si{\giga\electronvolt}$. It will also provide a platform for improved calibration of the existing detector. The enhanced understanding of deep ice optical properties will reduce the main systematic error that contributes to the directional uncertainty of astrophysical neutrinos.

An optical module is the basic building block of a large-volume neutrino telescope. Traditionally, it features a single large PMT measuring the amount of incoming photons (derived from the signal amplitude) as well as their arrival times. Novel optical sensors will play a key role in the expected performance enhancements of the IceCube Upgrade. A large fraction will be so-called multi-PMT Digital Optical Modules (mDOMs) featuring 24 relatively small (three-inch class) PMTs (see Sec.\ref{sec:pmts}). This multi-PMT approach, introduced to deep-sea detectors by the KM3NeT Collaboration \cite{Loehner2013:km3net_module}, results in attractive advantages with respect to the traditional single-PMT technique. The benefits include a homogeneous solid angle coverage and a large sensitive area per module. Furthermore, the mDOM provides not only the number and arrival time of photons, but also directional information as well as the possibility of multi-hit triggering inside one module (see Sec.~\ref{sec:calib_studies}).
%
\begin{figure}[tb]
\begin{minipage}{0.5\textwidth}
    \centering  
    \includegraphics[trim = 0cm 30cm 0cm 10cm, clip, width=0.6\textwidth]{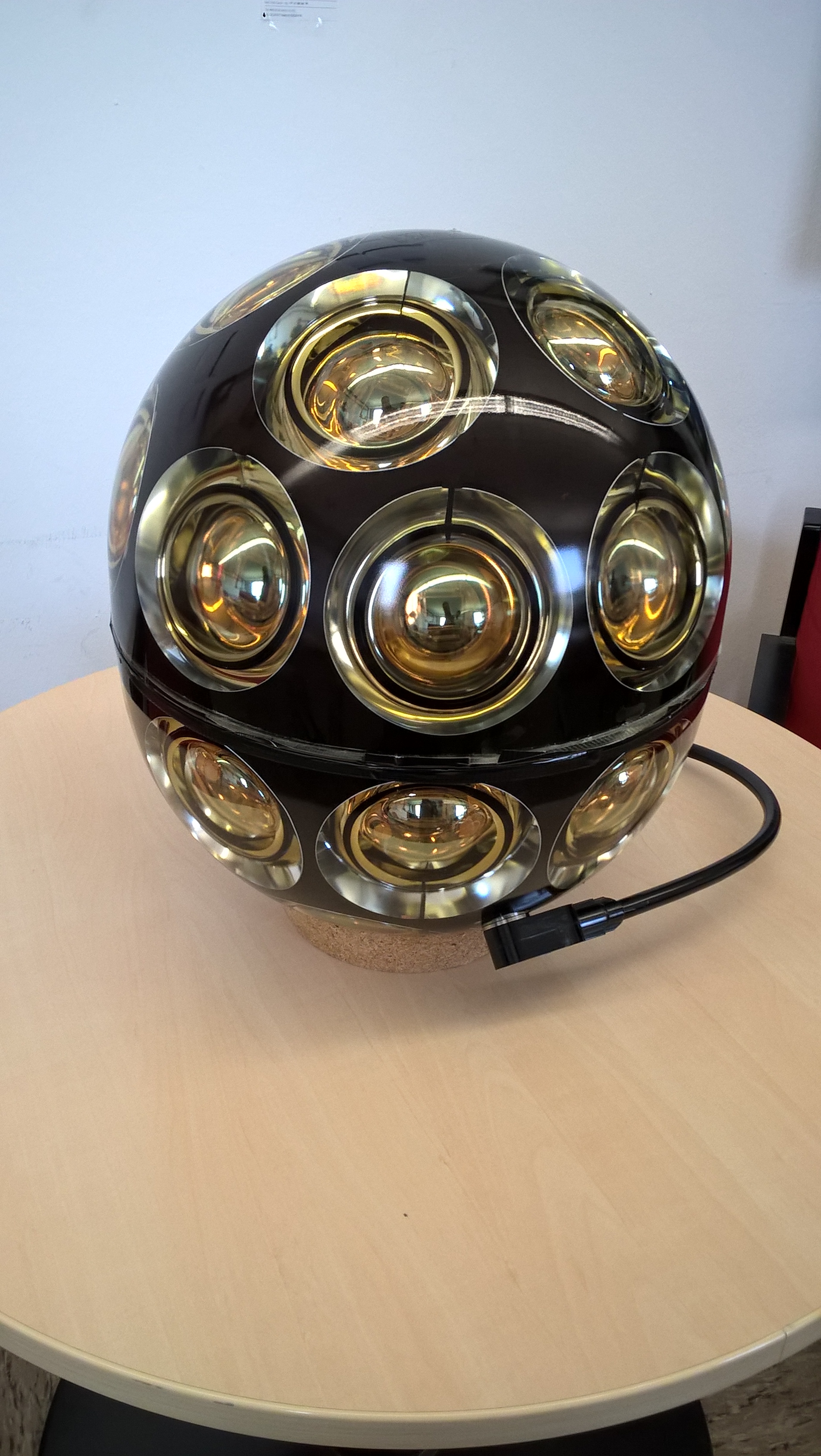}
\end{minipage}
\hfill
\begin{minipage}{0.5\textwidth}
    \centering
    \includegraphics[trim = 5cm 0cm 5cm 0cm, clip, width=0.9\textwidth]{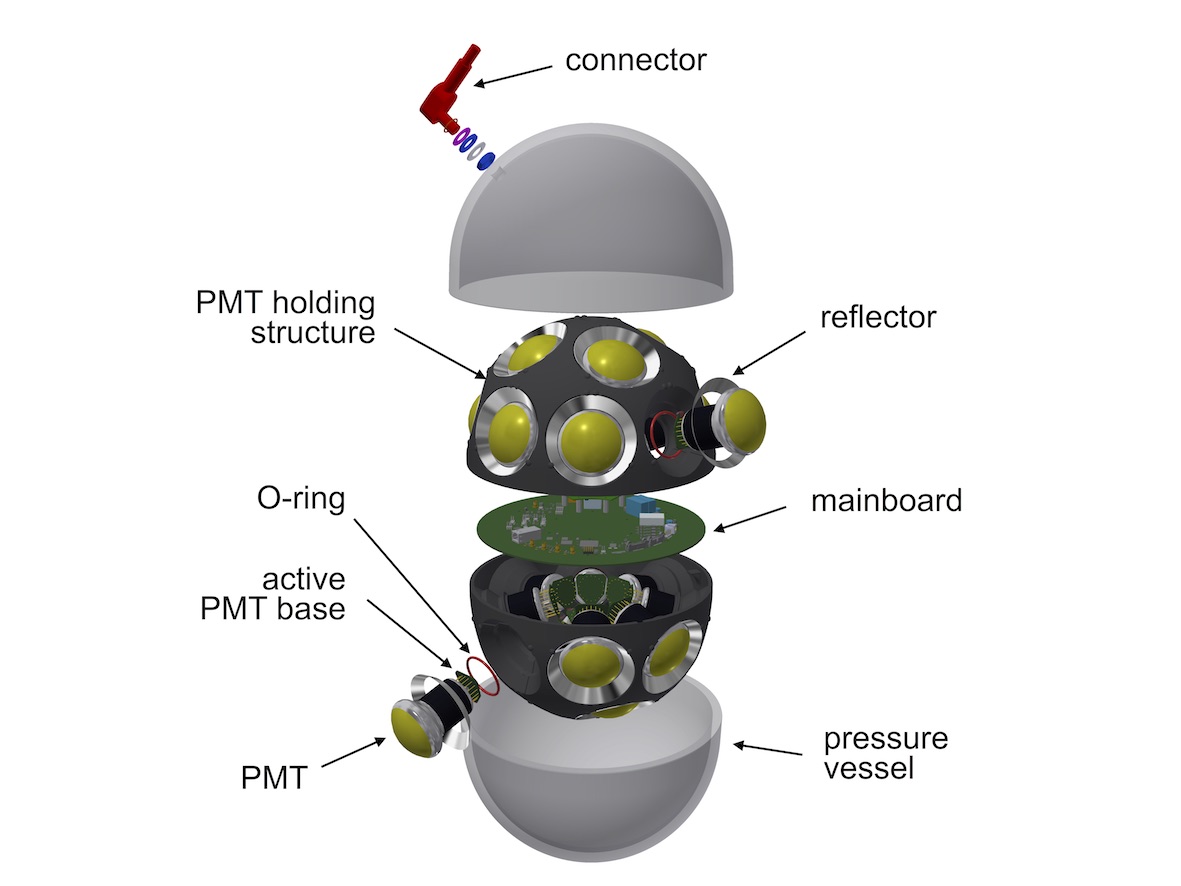}
\end{minipage}
\caption{mDOM overview: \emph{Left}: Demonstrator module (see Sec.~\ref{sec:status}) \emph{Right}: Exploded view featuring main components.}
\label{fig:mDOM}
\end{figure}
%


\section{Project Status}
\label{sec:status}
The deep-ice environment and detector infrastructure at the South Pole comes with unique challenges to optical module technology, such as a limited borehole diameter, pressure surges during freeze-in up to more than twice the static pressure and a tight power budget per module. The mDOM design (see Fig.~\ref{fig:mDOM}) was driven by these challenges. Below we present the design decisions and current status for key mDOM components.

\subsection{Photomultipliers and reflectors}
\label{sec:pmts}
Since our mDOM design is based on the KM3NeT multi-PMT module, the baseline PMT is a derivative of the well-characterized KM3NeT model from Hamamatsu. The mDOM, however, introduces stricter spacial constraints: the PMT, including the base, is limited to a conical volume with an opening angle of $\SI{40}{\degree}$ and a length of $\sim \SI{95}{mm}$.  

\begin{figure}[tb]
\centering
\includegraphics[trim=0 21cm 0 0, clip, width=\textwidth]{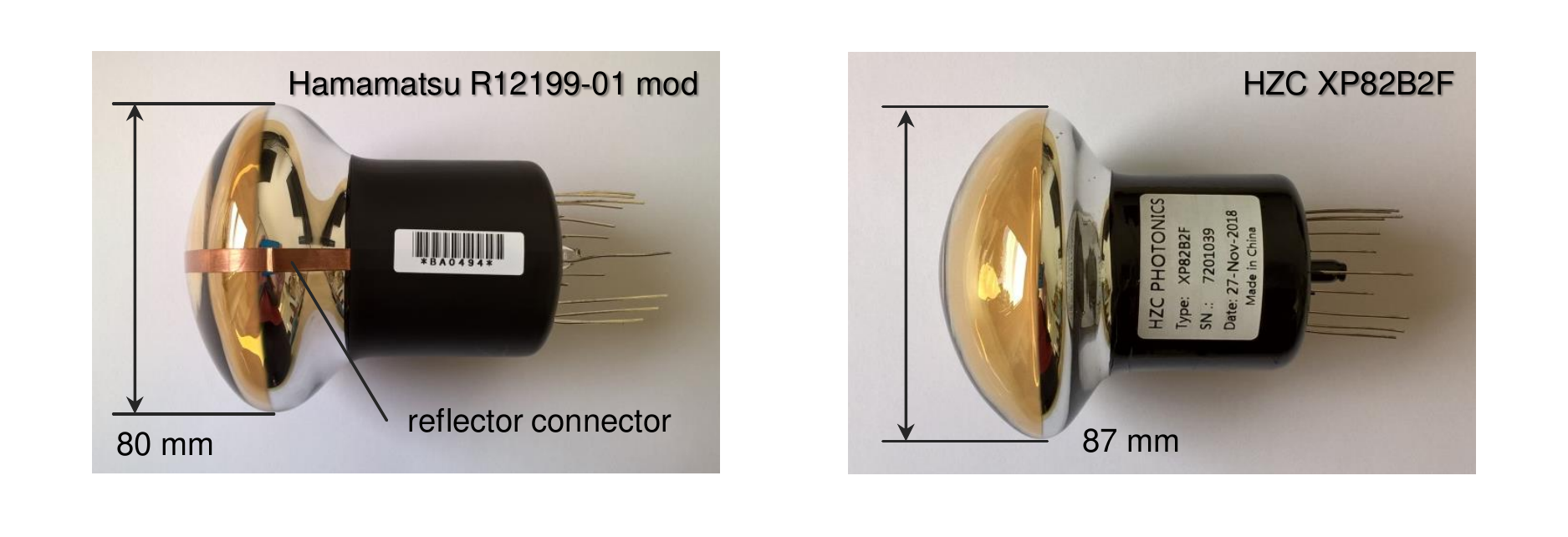}
\caption{Candidate PMTs for the multi-PMT Digital Optical Module (mDOM).}
\label{fig:pmts}
\end{figure}

The Hamamatsu type R12199-01 MOD HA PMT (Fig.~\ref{fig:pmts}, \textit{left}) is an 80-mm diameter photomultiplier based on the KM3NeT model R12199-02. It was modified for use in low-background applications with tight spacial constraints. To fit this usage profile, the PMT features a tube that has been reduced in length from $\SI{97}{mm}$ to $\SI{91}{mm}$.
An alternative PMT candidate is the model XP 82B2F produced by HZC photonics (see Fig.~\ref{fig:pmts}, \textit{right}) which features a larger tube diameter of $\SI{87}{\milli\meter}$. At the moment, detailed comparison studies of key parameters including gain slope, spurious pulsing probability, dark rate, quantum efficiency and timing are carried out in the IceCube collaboration \cite{icrc2019:pmts}. 

For easier signal readout in the mDOM, the PMTs will be operated with negative high voltage at the photocathode and a grounded anode (``negatively fed''). In order to reduce and stabilize the dark-noise rate evoked by this configuration both PMT models are equipped with a conductive layer located on the outside surface of the glass tube surrounding the electron multiplier system, which is electrically connected to the photocathode. 

Reflectors mounted around the entrance window of the PMTs increase the effective area for vertical (on-axis) illumination and compensate for photons lost due to shadowing of the support structure (see Sec.~\ref{sec:support_structure}) and absorption in glass and gel. The reflectors are produced from enhanced-reflectivity coated aluminum sheets. The material was chosen to maximize the overall reflectivity taking into account the spectrum of Cherenkov radiation, the quantum efficiency of the PMTs as well as the transparency of the pressure vessel glass. With respect to realistic alternative configurations the utilization of reflectors yields an extra $\sim 20\%$ in overall module sensitivity. In a Monte Carlo study simulating the incidence of a plane wave-front, the reflector opening angle was optimized to provide maximum sensitivity for vertical photons, focusing the PMT's field of view. This resulted in a final angle of $\SI{102}{\degree}$. The effective concentration of the photon acceptance at small incidence angles results in a loss of overall sensitivity $< 1$\% compared to the achievable maximum.

As conductive objects at non-photocathode potential close to the photocathode of a negatively fed PMT tend to evoke high and unstable dark-noise rates \cite{Unland2019:pmts} the reflectors will be connected to photocathode potential via dedicated copper strips (see left picture in Fig.~\ref{fig:pmts}).

\subsection{Electronics}
\label{sec:electronics}
The IceCube-Upgrade uses a modular architecture of common electronics components (communication, timing etc.) with well-defined interfaces to reduce the development efforts for the different optical module types. General requirements for the readout and high voltage systems are low power consumption, low sensitivity to interference signals, small dimensions and high reliability, paired with the ability to sample semi-complex waveforms from the PMTs.

The 24 PMTs of the mDOM are equipped with so-called ``active'' bases. These bases feature a low-power Cockcroft-Walton circuitry for individual in-situ high-voltage generation, allowing for the operation of all PMTs at identical gains. In the analog front-end, located on the main board in the center of the mDOM, the PMT pulse is amplified and then split. One path is fed into a comparator with a threshold of $\sim \SI{0.2}{pe}$. The comparator output is sampled with nanosecond resolution in a field-programmable gate array (FPGA) to precisely capture the rise time and time-over-threshold of the PMT pulse. It is also used to initiate the storage of the analog-to-digital converter (ADC) data. The signal on the second path is shaped to allow its continuous digitization with an ADC operated at a sampling frequency of $\SI{100}{\mega\hertz}$ located on the main board. The power consumption of the mDOM in this configuration is expected to be about $\SI{5}{\watt}$.
%

\begin{figure}[tb]
\begin{minipage}{0.5\textwidth}
  \centering
  \includegraphics[trim=2.5cm 0 2.5cm 0, clip, width=8cm]{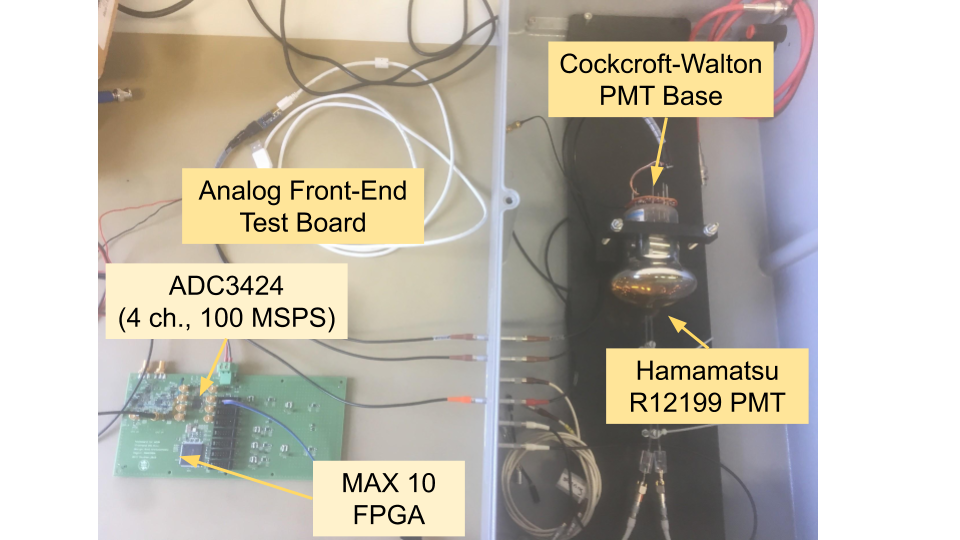}
\end{minipage}
\hfill
\begin{minipage}{0.5\textwidth}
  \centering
    \includegraphics[trim=2.5cm 0 2.5cm 0, clip, width=8cm]{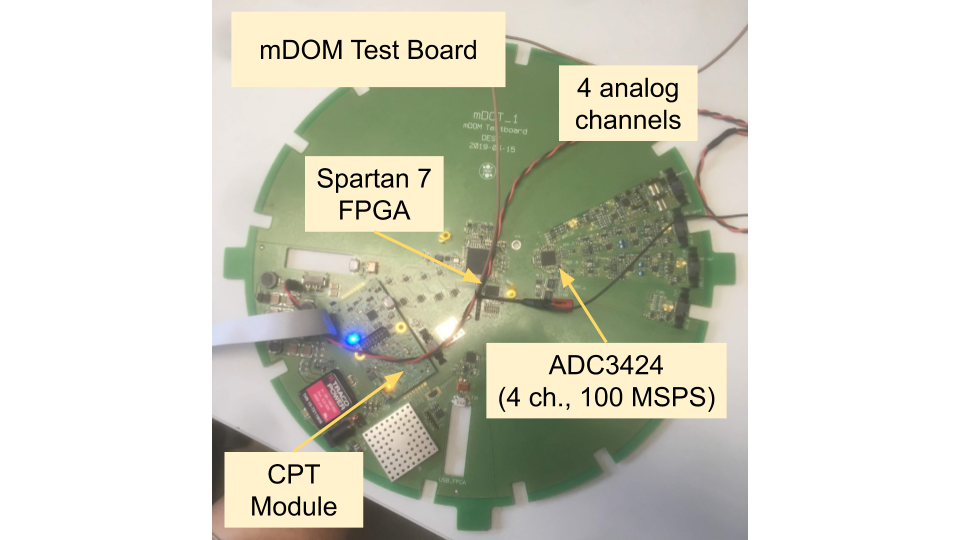}
\end{minipage}
\caption{\emph{Left}: Test setup featuring a Hamamatsu PMT equipped with an active base read-out by the analog front-end test board. \emph{Right}: mDOM Test Mainbaord (mDOT). The board features the final main board form factor and four different analog front-end designs.}
\label{fig:electronics}
\end{figure}

The test setup shown in Fig.~\ref{fig:electronics} (\textit{left}) comprises a PMT in a dark box connected to a prototype mDOM Cockcroft-Walton high voltage base. The PMT signal is processed in an analog front-end testboard comprising the comparator, pulse shaping and a $\SI{100}{\mega\hertz}$ ADC, read out by a MAX 10 FPGA allowing sampling and timestamping of the comparator output at 600 MHz. The setup is used to demonstrate the readout principle and optimize the analog signal path and pulse shaping. 

The mDOM Test Mainboard (mDOT) comprises the final form factor of the mainboard, the final FPGA and the Communication, Power, and Timing (CPT) module that will be used in all IceCube Upgrade in-ice devices. It has four different analog front-end designs for validation and testing. The final mainboard design will comprise a total of 24 analog-front channels (6 four-channel ADCs) mounted in ``pie-segments'' on the mainboard as can be seen in Fig.~\ref{fig:electronics} (\textit{right}). The ADCs and comparators are read out by a Spartan 7 FPGA in the center of the board. The final design will also add a microprocessor and 2 GB of DDR3 RAM for data processing and buffering.
The mDOT will enable us to perform tests with an integrated mDOM module, operating up to four PMTs at the same time, validating many of the physics requirements and conducting electromagnetic interference studies in a closed mDOM.

\subsection{Calibration devices}
\label{sec:calib_devices}
One of the main goals of the IceCube Upgrade is an enhanced understanding of the optical properties of the South Polar ice. A better understanding of the detector medium will reduce systematic uncertainties in the directional reconstruction of the neutrino enabling future high-precision measurements as well as a re-analysis of IceCube data taken in the past. To this end, it is foreseen to equip the mDOM with three cameras (including illumination devices) \cite{icrc2019:cameras}, ten LEDs, as well as acoustic sensors \cite{icrc2019:acoustics}. 

\subsection{Support structure and optical gel}
\label{sec:support_structure}
A support structure is used to position the internal components, that is the PMTs, reflectors, calibration devices and the main electronics board, inside the pressure vessel of the mDOM and seal the inner volume from the optical gel. The support structure is produced via selective laser sintering (a form of 3D printing) from white polyamide and subsequently died black with non-conductive color. Black was chosen because a white structure only marginally increases the overall module sensitivity while enhancing the amount of undesired photon back-scattering into the ice. An alternative material being investigated is glass-enhanced polyamide which features a $\sim$30\% reduced thermal expansion coefficient. This allows to reduce the stress acting on the optical gel layer during cooling of the module as a result of the different expansion coefficients of the glass vessel and the support structure.

Curing two-component silicone gel, occupying the space between the support structure, the PMTs and the pressure vessel, provides optical coupling\footnote{An mDOM configuration without optical gel coupling would result in an overall sensitivity loss of $\sim 55\%$.} as well as structural stability, cushioning, and electrical insulation between the individual PMTs. The baseline choice for prototyping is QGel 900 by Quantum Silicones which has been employed in the original IceCube optical module and proven to be suited for low ambient temperatures. As an alternative, Shin-Etsu 3539 gel is under consideration which is used in the D-Egg module \cite{icrc2019:degg}. With respect to QGel 900 it features a better optical transmission as well as a harder consistency.

\subsection{Pressure vessel}
\label{sec:vessel_gel}
\begin{figure}[tb]
\begin{minipage}{0.6\textwidth}
  \centering
    \includegraphics[width=0.9\textwidth]{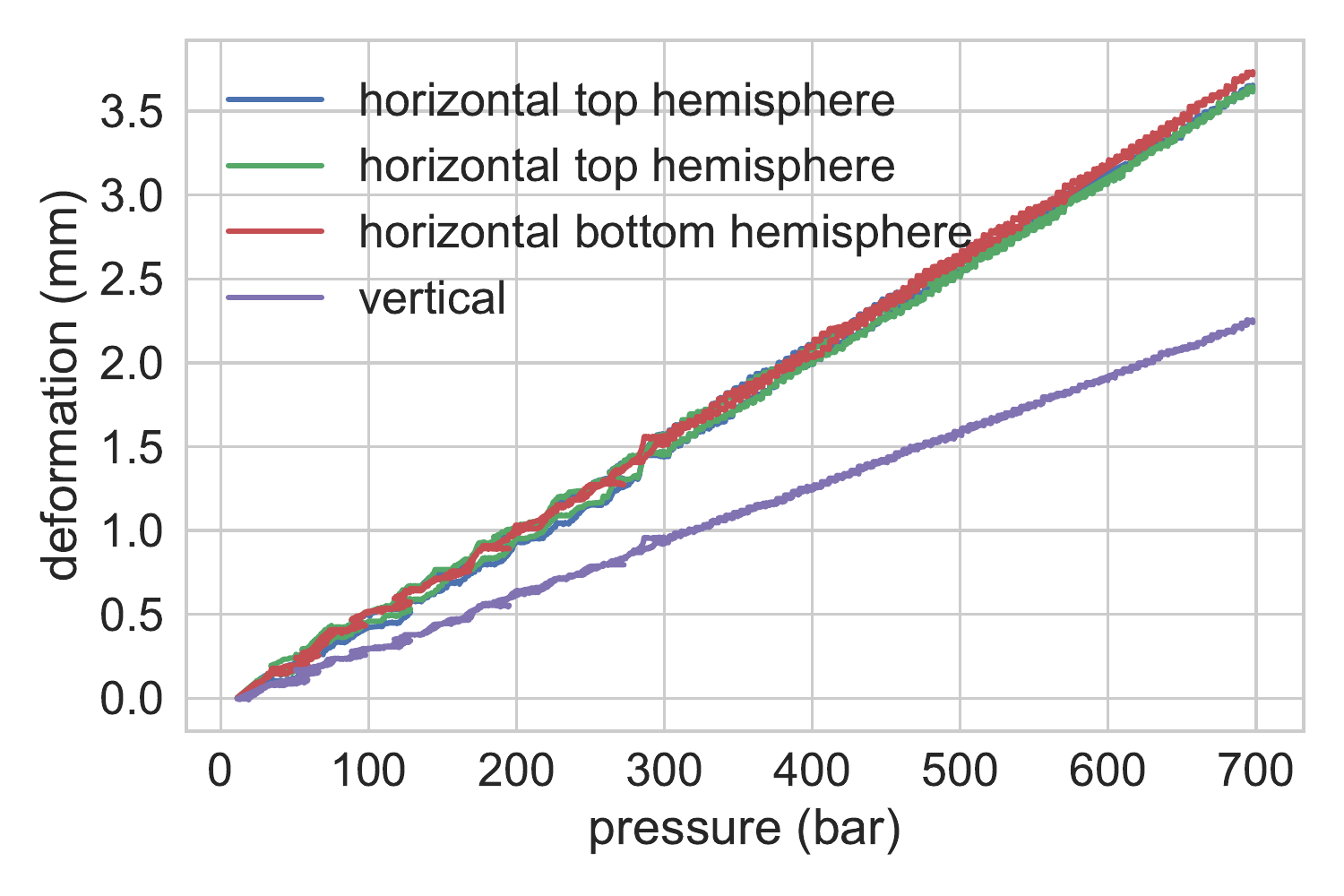}
\end{minipage}
\hfill
\begin{minipage}{0.4\textwidth}
  \centering
  \caption{Horizontal  and vertical (\textit{violet} line) shrinkage of an mDOM pressure vessel as function of external pressure in a hyperbaric chamber. The horizontal deformation was measured separately in the upper (two sensors, \textit{blue} and \textit{green} lines, partially hidden) and lower (\textit{red} line) vessel half.}
  \label{fig:pressure_test}
\end{minipage}
\end{figure}

With its cylindrical extension around the equator (see Fig.~\ref{fig:mDOM}), the mDOM pressure vessel slightly deviates from the spherical form of traditional optical modules. The design is driven by the limited diameter of the borehole and the non-reducible length of the PMTs. The mDOM vessel is rated for external pressures of $\SI{700}{\bar}$. This criterion, which is substantially higher than the hydrostatic pressure at the location of the detector, is based on pressure peaks  during re-freezing where up to $\SI{550}{\bar}$ were measured during IceCube construction. The pressure rating has been verified in hyperbaric tests: Fig.~\ref{fig:pressure_test} shows the reduction in horizontal and vertical diameter of the pressure vessel as function of external pressure. The shrinkage is reversible and agrees well with calculations based on finite element simulations. 

Optical modules will be connected to neighboring modules and the downhole cable via steel cables which are attached to the harness of the module. For the mDOM, two harness design options are being investigated: one is based on a mechanical frame enclosing the pressure vessel and the other on gluing stainless steel plates with mounting points for the cables to the pressure vessel.


\subsection{mDOM demonstrator}
In order to test and optimize the mDOM design, so-called demonstrator modules have been constructed. These modules feature key mechanical and optical components but no calibration devices. Their assembly showed the overall viability of the gel-filling process and the chemical compatibility of all components with the optical gel, but also revealed problems with the thermal expansion of the support structure during cooling to low temperatures. 
The demonstrator will provide a test-bed for the newly developed active bases (see Sec.~\ref{sec:pmts}) and the four-channel version of the mainboard (see Sec.~\ref{sec:electronics}).


\section{Sensitivity and Calibration Studies}
\label{sec:calib_studies}
Designed for the detection of high-energy neutrinos IceCube is also sensitive to low-energy ($\sim \mathrm{MeV}$) neutrinos from nearby supernovae. Supernova neutrinos are detected via a collective rise of the background rate in all optical modules. We studied the possibility to use local coincidences (multiple hits inside individual mDOMs) to suppress background noise and enhance the detection of supernovae. We modelled the neutrino flux from a type IIp supernova with a progenitor mass of 27 solar masses (using the LS220 equation of state model from \cite{Sukhbold2015:sn}) and simulated a future detector comprising 10\,000 mDOMs, similar to the possible design of IceCube-Gen2 \cite{Aartsen2014:gen2}, a considered high-energy extension of IceCube. In contrast to the current IceCube capability the technique was found to be sensitive to spectral parameters of the neutrino flux (mean energy and the so-called form parameter). The local coincidence trigger allows IceCube-Gen2 to detect supernovae up to a distance of $\sim \SI{300}{\kilo\parsec}$ with one false discovery per century while the IceCube detector reaches up to $\sim \SI{50}{\kilo\parsec}$ with one false discovery every ten years.
%
The study takes into account background contributions from the solar neutrino flux, the dark-noise rate of the PMTs and radioactive decays in the pressure vessel glass. Not considered are Michel electrons produced by atmospheric muons, correlated noise in an individual PMT as well as potential cross-talk between different PMT channels in the mDOM  \cite{Lozano2017:sn, Sprenger2019:sn}. 

A technique using background signals from radioactive decays inside the vessel glass for in-situ calibration was developed. The method allows to detect systematic offsets between individual PMTs on the $\SI{0.1}{\nano\second}$ level. The viability of the method was confirmed in first lab tests using a reduced mDOM configuration consisting of four PMTs \cite{Eder2019:calib}. 


\section{Conclusion and Outlook}
\label{sec:conclusion}
The multi-PMT Digital Optical Module (mDOM) is being developed for deployment in the deep ice at the South Pole for future IceCube extensions (IceCube Upgrade, IceCube-Gen2 \cite{Aartsen2014:gen2}). Infrastructure constraints and harsh environmental conditions pose stringent limits on the module parameters like size, power consumption and reliability. 

The mechanical design of the mDOM is well advanced and is currently being scrutinized and optimized. The Hamamatsu R12199-01 HA MOD PMT, a well-characterized model suited for the application, is considered the baseline model for the mDOM and was used in the construction of a demonstrator module. The HZC XP82B2F model could be an alternative with the potential to further enhance the mDOM sensitivity. For the readout and voltage generation, electronics baseline designs have been developed and prototypes are being evaluated. Two demonstrator modules have been built. Their construction provided valuable experience concerning technical details of the integration process and optimization potentials. These modules will be further utilized for heat transport tests, measurements of the angular acceptance and potential cross-talk between individual PMTs as well as tests of the new Cockcroft-Walton PMT base. New calibration and physics applications of the mDOM, such as supernova detection, are also being investigated.

Features specific to the mDOM (such as directional information and local coincidences) are expected to significantly enhance the energy and directional reconstruction capabilities of the IceCube Upgrade. Corresponding simulation studies to demonstrate this potential are currently under way.

\bibliographystyle{ICRC}
\bibliography{references}

%

\end{document}